\documentclass[floatfix,prb,aps,twocolumn,nobalancelastpage,raggedbottom,raggedfooter,showpacs]{revtex4}

\usepackage{graphicx,amssymb,amsmath}
\usepackage{dcolumn} 
\begin{document}

\title{Energy levels in polarization superlattices: a comparison of
  continuum strain models}

\author{B. Jogai}
\email[e-mail: ]{brahmanand.jogai@wpafb.af.mil} 
\affiliation{Air Force Research Laboratory, Materials
  and Manufacturing Directorate, Wright-Patterson Air Force Base, OH
   45433}
\affiliation{Semiconductor Research Center, Wright State University,
   Dayton, OH 45435}
\author{J. D. Albrecht}
\affiliation{Air Force Research Laboratory, Wright-Patterson Air Force
   Base, Ohio 45433}
\author{E. Pan}
\affiliation{Department of Civil Engineering, The University of Akron,
   Akron, Ohio 44325}

\begin{abstract}
  A theoretical model for the energy levels in polarization
  superlattices is presented.  The model includes the effect of strain
  on the local polarization-induced electric fields and the subsequent
  effect on the energy levels.  Two continuum strain models are
  contrasted.  One is the standard strain model derived from Hooke's
  law that is typically used to calculate energy levels
  in polarization superlattices and quantum wells.  The
  other is a fully-coupled strain model derived from the thermodynamic
  equation of state for piezoelectric materials.  The latter is more
  complete and applicable to strongly piezoelectric materials
  where corrections to the standard model are significant.
  The underlying theory has been applied to
  AlGaN/GaN superlattices and quantum wells.  It is found that the
  fully-coupled strain model yields very different electric fields
  from the standard model.  The calculated intersubband transition
  energies are shifted by approximately 5 -- 19 meV, depending on the structure.
  Thus from a device standpoint, the effect of applying the
  fully-coupled model produces a very measurable shift in the peak
  wavelength.  This result has implications for the design of
  AlGaN/GaN optical switches.
\end{abstract}


\maketitle 

\section{Introduction}
\label{sec:intro} Intersubband optical transitions (ISBT) in
AlGaN/GaN superlattices (SLs) and multiple quantum wells (MQWs)
are being exploited for use in near- and mid-infrared lasers and
ultra-fast all-optical switches in the 1.5 -- 3 $\mu$m wavelength
range.\cite{NSuzuki1,NSuzuki2,NSuzuki3,Iizuka1,Heber,Iizuka2,Kishino}
A key design issue related to ISBT-based device concepts is the
calculation of the electron energy levels of these structures so
that the peak wavelengths can be estimated before growth and
fabrication. In the present work, we investigate the role of
strain and polarization on the subband structure of SLs in the
wurtzite crystal structure. Through a theoretical examination of
fully-coupled and semi-coupled electromechanical treatments we
show the importance of using a fully-coupled model for predicting
the energy levels of SLs in strongly piezoelectric material
systems. The model is then used to predict the energy levels of
ten actual structures in order to compare our calculated ISBTs to
previously measured spectra.

Unlike zincblende semiconductor SLs, a number of issues arise in
wurtzite SLs that complicate the task of calculating the energy
levels.  AlGaN in the wurtzite phase has a large spontaneous
polarization moment along the $ [000\bar{1}] $ axis. In addition,
SLs grown on a SiC or sapphire substrate are pseudomorphic and the
large in-plane biaxial strains induce a piezoelectric polarization
moment oriented along the $c$-axis with the direction depending on
whether the strain is tensile (SiC) or compressive (Al$_2$0$_3$).
The discontinuity of the polarization moments effectively
represents fixed sheet charges at the interfaces of the SL.  In
general, each AlGaN on GaN interface in the direction $
[000\bar{1}] $ will have a positive space charge and each GaN on
AlGaN interface a negative space charge. Thus unlike zincblende
SLs which are flat band unless doped, a calculation of the
electronic eigenvalues using the Schr{\"o}dinger equation must be
preceded by a calculation of the electrostatic potential,
representing the Hartree term in the Schr{\"o}dinger equation,
using the Poisson equation.

There is a further complication that has been previously ignored
in calculations of the energies in AlGaN/GaN SLs.  This involves
the incorporation of strain into the electric field and eigenvalue
calculations.  To date, the strain model for AlGaN/GaN SLs has
been borrowed from the zincblende realm\cite{smith} without
additional consideration given to its validity for strongly
piezoelectric materials.  Although piezoelectric, zincblende
materials have comparatively small piezoelectric tensor elements
so that the thermodynamic equation of state is reduced to the
standard Hooke's law with little or no error.  From this relation,
the strain tensor for zincblende SLs can be readily worked out
with good accuracy. On the other hand, wurtzite materials have
large piezoelectric coefficients indicating strong coupling
between the strain and electric fields. In the present case of
group III-nitride materials, we will show that treating the
mechanical strain as separate from the electronic properties is no
longer a sound methodology. The result is that a linear
stress-strain model (Hooke's law) is no longer valid, and the
fully-coupled thermodynamic equation of state must be invoked to
obtain the strain and electric fields simultaneously.

The coupling described in the present work is somewhat analogous
to the electromechanical coupling in surface acoustic wave (SAW)
devices using AlN and GaN thin
films.\cite{Lobl,Naik,Cheng,Palacios,Takagaki} The strength of the
interaction between the electronic and mechanical properties in
SAW devices is determined by the electromechanical coupling
coefficient,\cite{Zelenka} a quantity that measures the
interaction between the acoustic and electromagnetic waves in
piezoelectric materials.\cite{Auld} In contrast, the
electromechanical coupling described herein deals with the
interaction between the \textit{static} electric and strain
fields.  Although the mathematical treatments of the two cases are
very different, both types of couplings originate from the same
thermodynamic equation of state.  In other work, the fully-coupled
theory has predicted deviations in the static strain fields
present in AlGaN/GaN heterostructure field-effect
transistors\cite{jogainew} and the idealized case of free-standing
(as opposed to substrate-conforming)
superlattices.\cite{freestand}

In previous modeling of piezoelectric SLs, the mechanical and
electronic properties are treated separately and sequentially:
\textit{(i)} first the in-plane strain is calculated from the
pseudomorphic boundary condition, \textit{(ii)} Hooke's law is
then invoked to obtain the longitudinal strain, and \textit{(iii)}
the calculated strain tensor is subsequently used as an input to
the Poisson and Schr\"odinger equations.  The strain is never
recalculated to reflect the presence of static electric fields in
the constituent layers of the SL. In this paper, we compare the
standard approach with a more rigorous continuum elastic theory
applicable to piezoelectric materials.  Using the proposed
formalism, we apply the fully-coupled equation of state for
piezoelectric materials to obtain simultaneously the strain and
electronic properties of AlGaN/GaN SLs.

First we treat the case of undoped SLs and show that closed-form
analytical expressions can be obtained for both the strain and
electric fields, following which the eigenstates can be calculated
using the Schr\"odinger equation.  It will be shown, using
specific examples, that the calculated strain and electric fields
differ substantially from those obtained using the standard
(uncoupled) strain model.  Depending on the Al fraction and the
geometry of the SL, the longitudinal strain calculated from the
standard model may be in error by as much as 40\% relative to the
fully-coupled model.  Further, it will be shown that the
calculated ISBT energy may differ from that of the standard model
by as much as 16meV, depending on the SL geometry.

Second, we treat the more useful case from a device standpoint of
SLs Si-doped in the well. (The doping can be tailored to populate
the lowest conduction subband to facilitate optical transitions.)
In this case, it is not possible to obtain closed-form analytical
expressions for the strain and electric fields. Instead, we use a
Schr\"odinger--Poisson solver in conjunction with the
fully-coupled equation of state.  The peak wavelength is
calculated for a number of structures and the results compared
with published experimental data. Once again it will be shown that
the standard and fully-coupled models yield significant
differences in the ISBT energy.

This paper is organized as follows: In Sec.\ \ref{sec:model} the
continuum strain model is described.  In Sec. \ref{subsec:slstrain},
the general equations for the fully-coupled strain model are obtained.
In Sec.\ \ref{subsec:poiseq}, the strain tensor and electric field for
a polarization SL are worked out.  In Sec.\ \ref{subsec:schrod}, the
calculation of the electron eigenstates is described.  A fully-coupled
numerical model is outlined in Sec.\ \ref{subsec:numerical}.
Calculated results are presented in Sec.\ \ref{sec:results}.  In Sec.\
\ref{subsec:undopedSL}, calculated results for the standard and
fully-coupled cases are contrasted for a model undoped SL.  In Sec.\
\ref{subsec:dopedSL}, both models are tested against published
experimental data for a series of doped SLs.  The results are
summarized in Sec.\ \ref{sec:summary}.

\section{Model description}
\label{sec:model} Ordinarily, calculating the strain or stress
tensor for a generalized strain problem becomes a complicated
numerical exercise involving minimizing the Helmholtz free energy
within the problem domain.\cite{Nye}  This approach suffices for
most materials, but specifically not for strongly piezoelectric
materials. The reason can be illustrated as follows: If we take a
piezoelectric plate and apply an external stress to it, the plate
will be geometrically deformed and, because of the piezoelectric
effect, a polarization moment will be induced, accompanied by an
internal electric field. But in addition to the piezoelectric
effect, there is also a \textit{converse} piezoelectric effect. In
our plate example, the induced electric field resulting from the
external stress will exert a counter force to resist deformation
of the plate. In a self-consistent way, the crystal will reach its
equilibrium state consonant with minimum stored energy. This
effect is present in all non-centrosymmetric crystals, but is
especially strong in certain hexagonal crystals. Consequently, the
uncoupled strain model for zincblende SLs results in errors when
applied to wurtzite SLs if the converse piezoelectric effect is
substantial, as in our case.

The relevant energy functional for piezoelectric materials is the
electric enthalpy $H$ given by\cite{ANSI}
\begin{equation}
  \label{eq:Hdef}
 H = U - \textbf{E}\cdot\textbf{D} ,
\end{equation}
where $\textbf{E}$ and $\textbf{D}$ are the electric field and
electric displacement, respectively, and $ U $ is the total
internal energy (strain + electrostatic) given by
\begin{equation}
  \label{eq:U}
  U = \frac{1}{2} C_{ijkl}
\gamma_{ij} \gamma_{kl} + \frac{1}{2} \varepsilon_{ij} E_i E_j,
\end{equation}
in which $ C_{ijkl} $ is the fourth-ranked elastic stiffness
tensor, $ \varepsilon_{ij} $ is the tensor form of the electric
permittivity, $ \gamma_{ij} $ is the strain tensor, and the
indices $i$, $j$, $k$, and $l$ run over the Cartesian coordinates
$x$, $y$, and $z$. Summation over repeated indices is implied
throughout.  Accompanying the energy functional is the
constitutive relationship for the electric displacement. For
piezoelectric materials, this is given, with the spontaneous
polarization included, by the expression
\begin{equation}
  \label{eq:Ddef}
  D_i = e_{ijk} \gamma_{jk} +\varepsilon_{ij} E_j+ P^\textrm{s}_i,
\end{equation}
in which $ e_{ijk} $ is the piezoelectric coefficient tensor and $
P^\textrm{s}_i $ is the spontaneous polarization.\cite{Bernardini}
For wurtzite materials, only the $z$ component of $ P^\textrm{s} $
exists because of the sixfold rotational symmetry of the [0001]
axis.  The first term in Eq.\ (\ref{eq:Ddef}) is the piezoelectric
polarization. After substitution into Eq.\ (\ref{eq:Hdef}), the
final form of the electric enthalpy becomes
\begin{equation}
  \label{eq:Hfinal}
  H = \frac{1}{2} C_{ijkl} \gamma_{ij} \gamma_{kl} -  e_{ijk}
  E_i \gamma_{jk} - \frac{1}{2} \varepsilon_{ij} E_i
  E_j - E_i P^\textrm{s}_i .
\end{equation}

\subsection{Fully-coupled strain tensor for planar strain}
\label{subsec:slstrain} In principle, minimizing $H$ within the
problem domain gives the strain and electric fields for a generalized problem.
In practice, this often means having to set up complicated finite
element calculations.  Problems involving two- and
three-dimensional geometric variations will be subjects of future
numerical work and we instead focus here on the SL case where the
issues of grid and minimization technique will not obscure the
physics. The SL problem is a planar one-dimensional (1-D) strain
problem with, at least nominally, no shear strains. Accordingly,
we can begin from the linear piezoelectric equation of state
\begin{equation}
  \label{eq:state}
  \sigma_{ij} = C_{ijkl} \gamma_{kl} - e_{kij} E_k ,
\end{equation}
obtained by differentiating Eq.\ (\ref{eq:Hfinal}) with respect to the
strain tensor, where $ \sigma_{ij} $ is the stress tensor.  Expanding
Eq.\ (\ref{eq:state}) and using the Voigt notation\cite{Nye} for the
third- and fourth-ranked tensors, the following stress-stain
relationships are obtained, assuming the $z$ axis to be the sixfold
axis of rotation:
\begin{subequations}
  \label{eq:allsig}
  \begin{equation}
    \sigma_{xx} = \gamma_{xx} C_{11} + \gamma_{yy} C_{12} +
    \gamma_{zz} C_{13} - e_{31} E_z ,
  \end{equation}
  \begin{equation}
    \sigma_{yy} = \gamma_{xx} C_{12} + \gamma_{yy} C_{11} +
    \gamma_{zz} C_{13} - e_{31} E_z ,
  \end{equation}
  \begin{equation}
    \label{subeq:sigzz}
    \sigma_{zz} = (\gamma_{xx} + \gamma_{yy} ) C_{13} + \gamma_{zz}
    C_{33} - e_{33} E_z ,
  \end{equation}
  \begin{equation}
    \sigma_{xy} = \gamma_{xy} ( C_{11} - C_{12} ) ,
  \end{equation}
  \begin{equation}
    \sigma_{xz} = 2 \gamma_{xz} C_{44} -  e_{15} E_x ,
  \end{equation}
and
  \begin{equation}
    \sigma_{yz} = 2 \gamma_{yz} C_{44} -  e_{15} E_y .
  \end{equation}
\end{subequations}
In the absence of the electric field, these equations are
recognized as the familiar tensor form of Hooke's law for
hexagonal crystals.

In conjunction with Eq.\ (\ref{eq:allsig}), we use the
constitutive relations obtained by expanding Eq.\ (\ref{eq:Ddef}):
\begin{subequations}
  \label{eq:Dall}
  \begin{equation}
    D_x = P_x + \varepsilon E_x ,
  \end{equation}
  \begin{equation}
    D_y = P_y + \varepsilon E_y ,
  \end{equation}
  \begin{equation}
    \label{subeq:Dz}
    D_z = P_z + \varepsilon E_z + P^\textrm{s} ,
  \end{equation}
\end{subequations}
where the electric permittivity is taken to be isotropic, a
reasonable approximation for AlGaN/GaN SLs, and the piezoelectric
moments are given by
\begin{subequations}
  \label{eq:Pall}
  \begin{equation}
    P_x = 2 e_{15} \gamma_{xz} ,
  \end{equation}
  \begin{equation}
    P_y = 2 e_{15} \gamma_{yz} ,
  \end{equation}
and
  \begin{equation}
    \label{subeq:Pz}
    P_z = e_{31} ( \gamma_{xx} + \gamma_{yy} ) + e_{33} \gamma_{zz} .
  \end{equation}
\end{subequations}

For simplicity, it is assumed that there are no shear strains,
manifested by warping, within the structure. The boundary
condition for a free surface, $ \sigma_{iz} = 0 $, can then be
applied throughout the layers, instead of just at the surface.
From Eq.\ (\ref{subeq:sigzz}), this gives
\begin{equation}
  \label{eq:ezz}
  \gamma_{zz} = -\frac{2 C_{13} }{C_{33}} \gamma_{xx} + \frac{e_{33}}{C_{33}}
  E_z ,
\end{equation}
where $ \gamma_{yy} = \gamma_{xx} $ in the 1-D planar case and $
\gamma_{xx} $ is assumed to be known from the pseudomorphic
condition across the interfaces.  There still remains the problem
of finding the electric field which is the topic of the next
section.

\subsection{Poisson equation}
\label{subsec:poiseq}
For an isolated piezoelectric plate under planar stress, the
constitutive equations and the equations of state should be sufficient
for obtaining the strain and electric fields.  For the SL, however,
the continuity of the electric displacement must satisfied at the
interface, and periodic boundary conditions must be imposed on the
electrostatic potential $ \phi $, as well as the continuity of $ \phi
$ across the interface.  Additional complications will arise from
doping, as this will give rise to space charges and free electrons.
These requirements are all met by solving the Poisson equation.  From
Gauss's law and Eqs.\ (\ref{subeq:Dz}), (\ref{subeq:Pz}), and
(\ref{eq:ezz}), we obtain the 1-D Poisson equation
\begin{eqnarray}
  \label{eq:pois1d}
  \frac{\partial}{\partial z} \left (  \kappa \frac{\partial \phi}{\partial
  z} \right ) = -e (N_d^+ - n) + \frac{\partial P^\textrm{s}}{\partial
  z} + \nonumber \\ 2
  \frac{\partial}{\partial z} \left [ \left ( e_{31} - e_{33}
  \frac{C_{13}}{C_{33}}
  \right ) \gamma_{xx} \right ] ,
\end{eqnarray}
where $ N_d^+ $ is the ionized donor concentration, $n$ is the free
electron concentration calculated from the Fermi energy and the wave
functions, $e$ is the electronic charge, and
\begin{equation}
  \label{eq:kappadef}
  \kappa = \varepsilon + \frac{e_{33}^2}{C_{33}} .
\end{equation}
It is evident from Eq.\ (\ref{eq:pois1d}) that $ \kappa $ serves
as an effective electric permittivity in the fully-coupled case.
Also, because $ e^2_{33} / C_{33} > 0 $, the electromechanical
coupling results effectively in additional dielectric screening.

Figure \ref{fig:superlat} shows the band edges for a period of the
SL under consideration.  In the following derivation, ``$a$''
refers to the AlGaN layer and ``$b$'' the GaN layer.  Equation
(\ref{eq:pois1d}) is solved to obtain $\phi$ subject to the
continuity of $ \phi $ and the electric displacement $ D_z $ at $
z = w_a $.  The latter is expressed by
\begin{equation}
  \label{eq:Dcontinue}
  \left . \kappa \frac{\partial \phi}{\partial z}
  \right |^{w_a^+}_{w_a^-} = \left . P^\textrm{s} \right
  |^{w_a^+}_{w_a^-} + \left . 2 \left ( e_{31} - e_{33}
  \frac{C_{13}}{C_{33}}  \right ) \gamma_{xx}  \right |^{w_a^+}_{w_a^-}  .
\end{equation}
It is assumed that there is no applied bias.  Periodic boundary
conditions then apply. This is accomplished by setting $ \phi = 0
$ at $ z = 0 $ and $ z = (w_a + w_b) $. Unless the free electron
distribution can be realistically approximated by a $\delta$
function, Eq.\ (\ref{eq:pois1d}) has to be solved numerically in
the most general case.  To illustrate the concept of
electromechanical coupling, we assume for the moment that the SL
is nominally undoped and with no free electrons from traps or
surface states.  Later, we will present results for doped SLs
using our fully-coupled numerical model.  For a depleted SL, the
general solution of Eq.\ (\ref{eq:pois1d}) is given by
\begin{equation}
  \label{eq:general}
  \phi =  \frac{P^\textrm{s}}{\kappa} z +
  \frac{2 (e_{31} C_{33} - e_{33} C_{13}) \gamma_{xx}}{\kappa
  C_{33}} z + \frac{A}{\kappa} z + B ,
\end{equation}
where $A$ and $B$ are unknown constants.  Thus there are four
unknowns, two in each layer.  All four constants are accounted for by
the four boundary conditions discussed above.

After obtaining the unknowns, the electric fields in the two layers
are given by
\begin{eqnarray}
  \label{eq:Fa}
  E_z^a = \frac{w_b ( P^\textrm{s(b)} - P^\textrm{s(a)} )}{w_a \kappa_b
  + w_b \kappa_a} + \frac{2 w_b ( e_{33}^\textrm{a} C_{13}^\textrm{a}
  - e_{31}^\textrm{a} C_{33}^\textrm{a} )
  \gamma_{xx}^\textrm{a}}{C_{33}^\textrm{a} (w_a
  \kappa_b + w_b \kappa_a )} \nonumber \\
  - \frac{2 w_b ( e_{33}^\textrm{b} C_{13}^\textrm{b} -
  e_{31}^\textrm{b} C_{33}^\textrm{b} )
  \gamma_{xx}^\textrm{b}}{C_{33}^\textrm{b} (w_a
  \kappa_b + w_b \kappa_a )} = - \frac{w_b}{w_a} E_z^b .
\end{eqnarray}
In the standard model, $ E_z^a $ and $ E_z^b $ are obtained by
replacing $ \kappa $ by $ \varepsilon $ in Eq.\ (\ref{eq:Fa}), using
the appropriate subscripts for the two layers.  It is seen, therefore,
that the fully-coupled electric field is smaller than its standard
counterpart.  From Eq.\ (\ref{eq:ezz}), the longitudinal strain in the
``$a$'' layer is given by
\begin{eqnarray}
  \label{eq:gammazza}
  \gamma_{zz}^\textrm{a} = - \frac{2
  C_{13}^\textrm{a}}{C_{33}^\textrm{a}} \gamma_{xx}^\textrm{a} +
  \frac{2 w_b e_{33}^\textrm{a} (e_{33}^\textrm{a} C_{13}^\textrm{a} -
  e_{31}^\textrm{a} C_{33}^\textrm{a}) \gamma_{xx}^\textrm{a}
  }{{C_{33}^\textrm{a}}^2 ( w_a \kappa_b + w_b \kappa_a ) }
  \nonumber  - \\
   \frac{2 w_b e_{33}^\textrm{a} (e_{33}^\textrm{b} C_{13}^\textrm{b}
  - e_{31}^\textrm{b} C_{33}^\textrm{b} )
  \gamma_{xx}^\textrm{b}}{C_{33}^\textrm{a} C_{33}^\textrm{b} ( w_a
  \kappa_b + w_b \kappa_a )} + \frac{w_b
  e_{33}^\textrm{a} (P^\textrm{s(b)} -
  P^\textrm{s(a)})}{C_{33}^\textrm{a} ( w_a \kappa_b + w_b \kappa_a )
  } ,
\end{eqnarray}
and in the ``$b$'' layer by
\begin{eqnarray}
  \label{eq:gammazzb}
  \gamma_{zz}^\textrm{b} = - \frac{2
  C_{13}^\textrm{b}}{C_{33}^\textrm{b}} \gamma_{xx}^\textrm{b} +
  \frac{2w_a e_{33}^\textrm{b} ( e_{33}^\textrm{b} C_{13}^\textrm{b} -
  e_{31}^\textrm{b} C_{33}^\textrm{b})
  \gamma_{xx}^\textrm{b}}{{C_{33}^\textrm{b}}^2 (w_a \kappa_b + w_b
  \kappa_a)}  - \nonumber \\
  \frac{2 w_a e_{33}^\textrm{b} (e_{33}^\textrm{a} C_{13}^\textrm{a} -
  e_{31}^\textrm{a} C_{33}^\textrm{a} )
  \gamma_{xx}^\textrm{a}}{C_{33}^\textrm{a} C_{33}^\textrm{b} (w_a
  \kappa_b + w_b \kappa_a)} - \frac{w_a e_{33}^\textrm{b}
  (P^\textrm{s(b)} - P^\textrm{s(a)})}{C_{33}^\textrm{b}(w_a
  \kappa_b + w_b \kappa_a)} .
\end{eqnarray}
We can compare these expressions for strain in the wurtzite system
directly with the zincblende case where the spontaneous
polarization terms vanish and the compliance tensor has fewer
unique elements. There, a similar (but somewhat less complicated)
expression to those in Eqs. (\ref{eq:gammazza}) and
(\ref{eq:gammazzb}) is obtained for the longitudinal strain in a
[111]-oriented pseudomorphic layer. The zincblende [111] case was
derived seperately by Bahder\cite{Bahder} using the method of
Lagrange multipliers to minimize the free energy density, an
alternate approach. In other work on lattice dynamics in undoped
GaN/AlN SLs,\cite{Gleize} comparable electric field corrections to
the strain along the growth direction are obtained with the main
difference being the use of the high-frequency dielectric
permittivity $\varepsilon (\infty)$ as opposed to the present case
of static screening, as in Eq. (\ref{eq:kappadef}).

The in-plane strains are calculated by assuming perfect in-plane
atomic registry of the SL layers with the buffer layer.  Applying this
condition, the in-plane strains are given by
\begin{equation}
  \label{eq:exxa}
  \gamma_{xx}^\textrm{a} = \frac{a_\textrm{bfr} - a_a}{a_a} ,
\end{equation}
and
\begin{equation}
  \label{eq:exxb}
  \gamma_{xx}^\textrm{b} = \frac{a_\textrm{bfr} - a_b}{a_b} ,
\end{equation}
where $a_a$ and $a_b$ are the relaxed $c$-plane lattice constants
of the ``$a$'' and ``$b$'' layers, respectively, and $
a_\textrm{bfr} $ is the $c$-plane lattice constant of the buffer
layer.  The foregoing model also works for less than perfect
registry: if the in-plane strains are known independently, they
can still be substituted into the above equations to obtain the
electric fields and longitudinal strains.  As is well known, the
standard model gives the longitudinal strains as
\begin{equation}
  \label{eq:gammazzastd}
  \gamma_{zz}^\textrm{a (std)} = - \frac{2
  C_{13}^\textrm{a}}{C_{33}^\textrm{a} } \gamma_{xx}^\textrm{a} ,
\end{equation}
and
\begin{equation}
  \label{eq:gammazzbstd}
  \gamma_{zz}^\textrm{b (std)} = - \frac{2
  C_{13}^\textrm{b}}{C_{33}^\textrm{b} } \gamma_{xx}^\textrm{b} ,
\end{equation}
i.e. the first terms in Eqs.\ (\ref{eq:gammazza}) and
(\ref{eq:gammazzb}), and, as a consequence, omits a great deal of
physics under certain conditions.  It will be seen shortly that
the fully-coupled correction to the standard longitudinal strain
is quite significant.

\subsection{Schr\"odinger equation}
\label{subsec:schrod} Owing to the large band gaps of the
consituents of the AlGaN/GaN SL, the electron eigenstates states
can be described by a Hamiltonian in the $ \Gamma_{7c} $ basis
without including any mixing from the $ \Gamma_{9v} $ and $
\Gamma_{7v} $ hole states, incurring little error in the process.
The resulting Hamiltonian is the one-band Schr\"odinger equation
\begin{eqnarray}
  \label{eq:schro_def}
  - \frac{\hbar^2}{2} \frac{\partial}{\partial z} \left (
      \frac{1}{m^* }
    \frac{\partial \Psi}{\partial z} \right ) + \frac{\hbar^2 (k_x^2 +
    k_y^2)}{2m^*} \Psi + \nonumber \\
    ( \Delta E_c - e \phi - e \phi_{xc} ) \Psi +
  a_c ( \gamma_{xx} + \gamma_{yy} +
    \gamma_{zz}  ) \Psi = E ( \textbf{k} ) \Psi , \nonumber \\
\end{eqnarray}
where $k_x$ and $k_y$ are the electron wave vectors in the $c$-plane,
$ \Psi $ is the electron wave function, $ E ( \textbf{k} ) $ is the
total electron energy, $ \phi $ is the electrostatic potential
discussed in Sec.\ \ref{subsec:poiseq} and represents the Hartree part
of the Coulomb interaction, $ \phi_{xc} $ represents the
exchange-correlation part of the Coulomb interaction, $ m^* $ is the
effective electron mass, $ a_c $ is the conduction band hydrostatic
deformation potential, $ \Delta E_c $ is the conduction band
discontinuity before strain shown in Fig.\ \ref{fig:superlat}, and $
\gamma_{ii} $ has been defined previously.

It should be noted that all of these quantities depend on $z$. For
$ \Delta E_c $, we assume that 60\% of the band gap difference
between the two materials appears in the conduction band, with the
caveat that the offset is not well known.  One could legitimately
use the conduction band offset as an adjustable parameter to try
to fit published experimental data, but it has been kept fixed in
the calculated results presented here.  There is a net hydrostatic
component of the strain obtained from the sum of the diagonal
elements of the strain tensor.  This component will shift the band
edge to higher or lower energy, depending on whether the
hydrostatic component is compressive or tensile.

If the SL is undoped, the electric field is piecewise constant so
that $ \phi = - F z $ in Eq.\ (\ref{eq:schro_def}) in the
respective layers.  Analytic solutions of the wave function then
can be obtained using Airy functions.\cite{Ridley2} A much more
flexible approach, however, and the one adopted in the present
work, is to use a discretized numerical technique, e.g.
finite-differencing, that can also handle the more technologically
interesting case of doped SLs. For SLs and MQWs, Bloch boundary
conditions are enforced, i.e. $ \Psi (0) = \Psi(w_a + w_b) \exp[
-i k_z (w_a + w_b) ] $, where $ k_z $ is the crystal momentum
corresponding to the periodicity of the layers along the growth
axis.

\subsection{Fully-coupled numerical model}
\label{subsec:numerical} For doped SLs in which the electrostatic
potential is very non-linear, the Poisson and Schr\"odinger
equations cannot be solved analytically in closed form. For this
case, we use a fully-coupled numerical model.  The central
framework for this is a Schr\"odinger--Poisson solver.  The
electron states and the free electron distribution are calculated
by solving Eq.\ (\ref{eq:schro_def}) on a finite-difference grid
subject to the boundary conditions discussed above.  If present,
hole states are calculated using a $ 6 \times 6 $ $ \textbf{k}
\cdot \textbf{p} $ Hamiltonian.  For the exchange-correlation
potential, we use the parameterized expression of Hedin and
Lundqvist,\cite{Hedin} derived from density-functional theory
within the local-density approximation. The charge-balance
equation, which determines the position of the Fermi energy $E_F$
in relation to the SL subbands, is solved by the Newton-Raphson
method.  Fermi-Dirac statistics are used for the probability of
occupancy of the electron states.

The model has been described in detail in Ref.\
\onlinecite{hfetpara} and the band structure and strain parameters
provided therein.  Since then, a fully-coupled strain calculation
has been added to the numerical model by solving the modified
Poisson equation, i.e. using $ \kappa $ instead of $ \varepsilon $
as shown in Eq.\ (\ref{eq:pois1d}), and incorporating Eq.\
(\ref{eq:ezz}) into the self-consistent calculation.  This means
that the strain terms in Eq.\ (\ref{eq:schro_def}) are updated
each time it is solved.  In an uncoupled calculation, the strain
terms would remain invariant throughout the self-consistent
calculation.  It has been shown\cite{Bernardini2} that there is a
bowing of the spontaneous polarization as a function of $x$.  This
effect is included in the present model.

\section{Results and discussion}
\label{sec:results}
First we show calculated longitudinal strains and electric fields for
a model undoped SL to illustrate the differences between the standard
and fully-coupled strain models.  We then show how these differences
lead to differences in the calculated eigenstates of the SL.  We then
present calculated ISBT energies and peak wavelengths using both the
standard and fully-coupled models for doped SLs and compare the
results with published experimental data.

\subsection{Undoped Superlattices}
\label{subsec:undopedSL} 
\begin{figure}
  \begin{center}
    \includegraphics[width=3.2in]{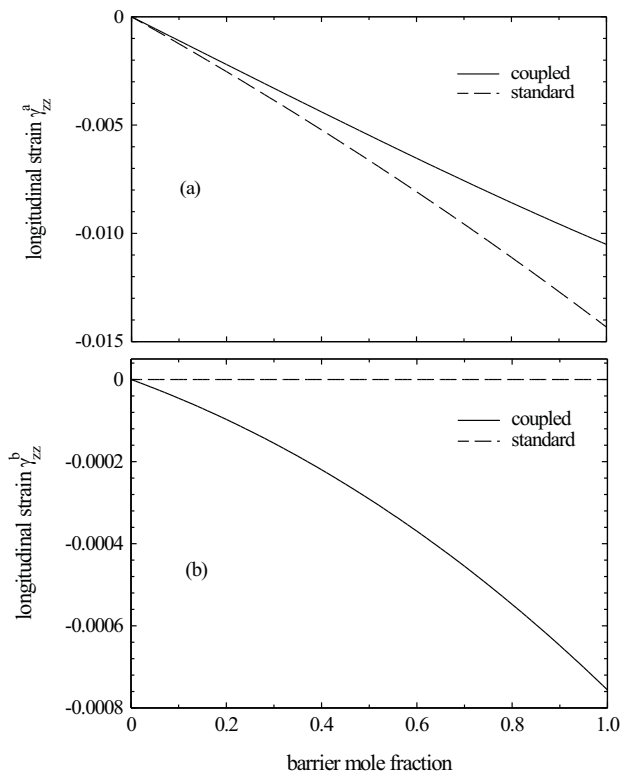}
  \end{center}
  \caption{Calculated longitudinal strain in (a) the barrier layer and
    (b) the well layer for an undoped SL consisting of 20\AA{} $
    \textrm{Al}_x \textrm{Ga}_{1-x} $N and 60\AA{} GaN on a GaN
    buffer as a function of $x$.  The fully coupled and standard
    results are shown.}
  \label{fig:ezz}
\end{figure}

We consider a model SL consisting of
20\AA{} $ \textrm{Al}_x \textrm{Ga}_{1-x} $N barriers and 60\AA{}
GaN wells on a GaN buffer. Assuming the pseudomorphic condition to
hold, the GaN layers will have no in-plane strain components,
while the AlGaN layers will have in-plane strains in accordance
with Eq.\ (\ref{eq:exxa}).  Using Eqs.\ (\ref{eq:Fa}),
(\ref{eq:gammazza}), and (\ref{eq:gammazzb}), the strain and
electric fields are calculated for the fully-coupled model and
compared with the standard results.  Following established
convention, a negative sign in the present calculations indicates
contraction and a positive sign extension relative to the
unstrained state.  Figures \ref{fig:ezz}(a) and (b) show the
longitudinal strain in the barrier and well layers, respectively,
as a function of the barrier mole fraction for the fully-coupled
and standard cases. Despite $\gamma_{xx}^\textrm{b}$$=$$0$ due to
the lattice matching condition, a non-zero $
\gamma_{zz}^\textrm{b} $ occurs due to electromechanical coupling,
as predicted in Eq.\ (\ref{eq:ezz}) and again in Eq.\
(\ref{eq:gammazzb}), and shown in Fig. \ref{fig:ezz}(b).  This
strain is a near linear function of the Al fraction and, in this
example, is about $ -0.07 $\% for $x$$=$$1$.

\begin{figure}
  \begin{center}
    \includegraphics[width=3.2in]{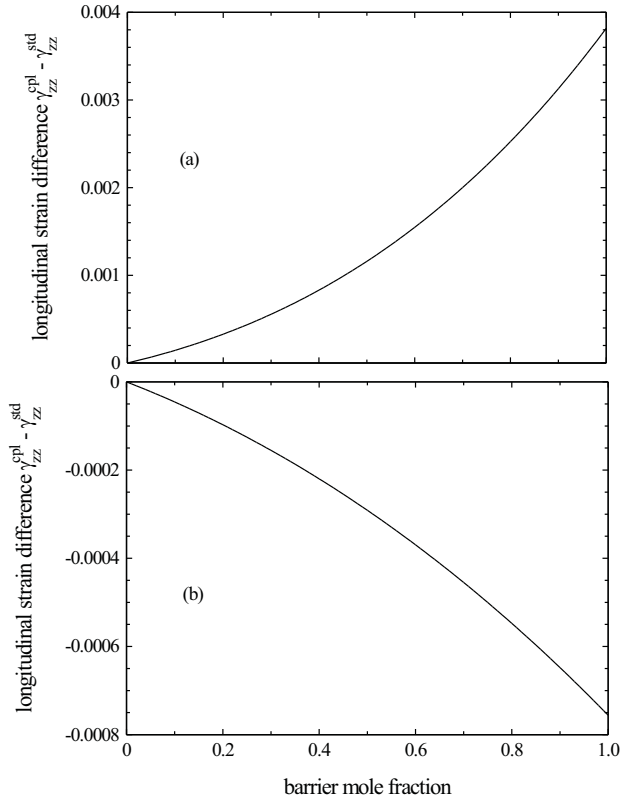}
  \end{center}
  \caption{Difference between  the fully  coupled and uncoupled
    longitudinal strains in (a) the barrier layer and (b) the well
    layer for the SL of Fig.\ \protect\ref{fig:ezz} as a function
    of Al composition $x$ in the barrier.}
  \label{fig:diffezz}
\end{figure}

The largest strains occur in the AlGaN layers due to the lattice
mismatch and it is also here that a significant deviation between
the standard and fully-coupled models is seen as shown in Figs.\
\ref{fig:ezz}(a) and (b).  This deviation is shown in Figs.\
\ref{fig:diffezz}(a) and (b). The error of the standard model
relative to the fully-coupled model can be in excess of 35\%
AlN/GaN SLs, as seen in this example.  It is even higher in
structures with higher electric fields in the barrier.  This
occurs when $w_b$$\gg$$w_a $. For example, if we set $w_a$=10\AA{}
and $w_b$=60\AA{}, the error for $ x = 1 $ is about 45\%. These
deviations in the strain are quite significant and, as will be
seen shortly, have an impact on the ISBT energies.

\begin{figure}
  \begin{center}
    \includegraphics[width=3.2in]{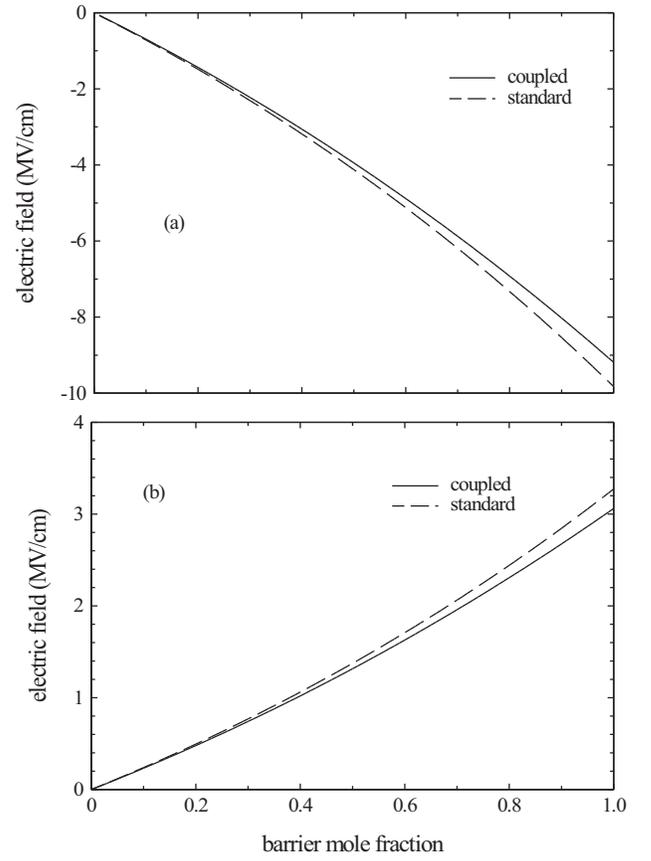}
  \end{center}
  \caption{Calculated electric field in (a) the barrier and (b) the
    well layers for the SL described in Fig.\ \protect\ref{fig:ezz}
    as a function of Al composition $x$ in the barrier.  The fully coupled (solid
    lines) and standard (dashed lines) results are shown.}
  \label{fig:efield}
\end{figure}

Figure \ref{fig:efield} shows the calculated electric fields in
the AlGaN and GaN layers for our model SL. From Eq.\
(\ref{eq:Fa}), it is seen that the larger electric field occurs in
the thinner layer.  The electric field calculated from the
fully-coupled model is smaller in magnitude than the standard
electric field due to an effective screening caused by the
electromechanical coupling.  This screening increases at higher
strains.  The deviation between the standard and fully-coupled
electric fields is about 7\% for $ x = 1 $.

\begin{figure}
  \begin{center}
    \includegraphics[width=3.2in]{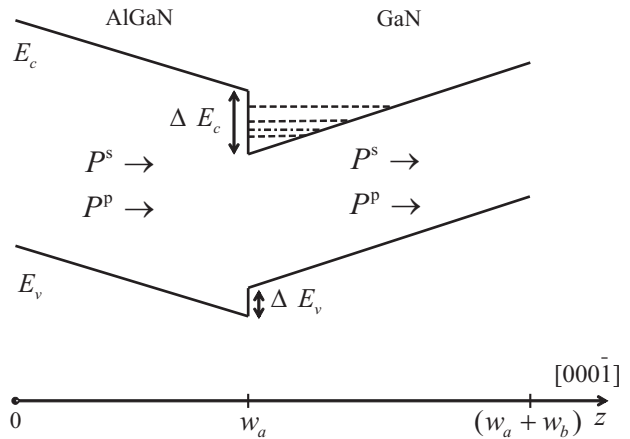}
  \end{center}
  \caption{Conduction and valence band edges of one period of a
    superlattice or MQW.  Assuming that the substrate is to the right,
    the schematic depicts a cation-faced structure.  $w_a$ and $w_b$
    are the thicknesses of the AlGaN and GaN layers, respectively.
    The directions of the spontaneous and piezoelectric moments in the
    two layers are indicated, assuming the buffer is GaN.  The dashed
    lines indicate the first three minibands and the dot-dashed line
    the Fermi energy.}
  \label{fig:superlat}
\end{figure}

\begin{figure}
  \begin{center}
    \includegraphics[width=3.2in]{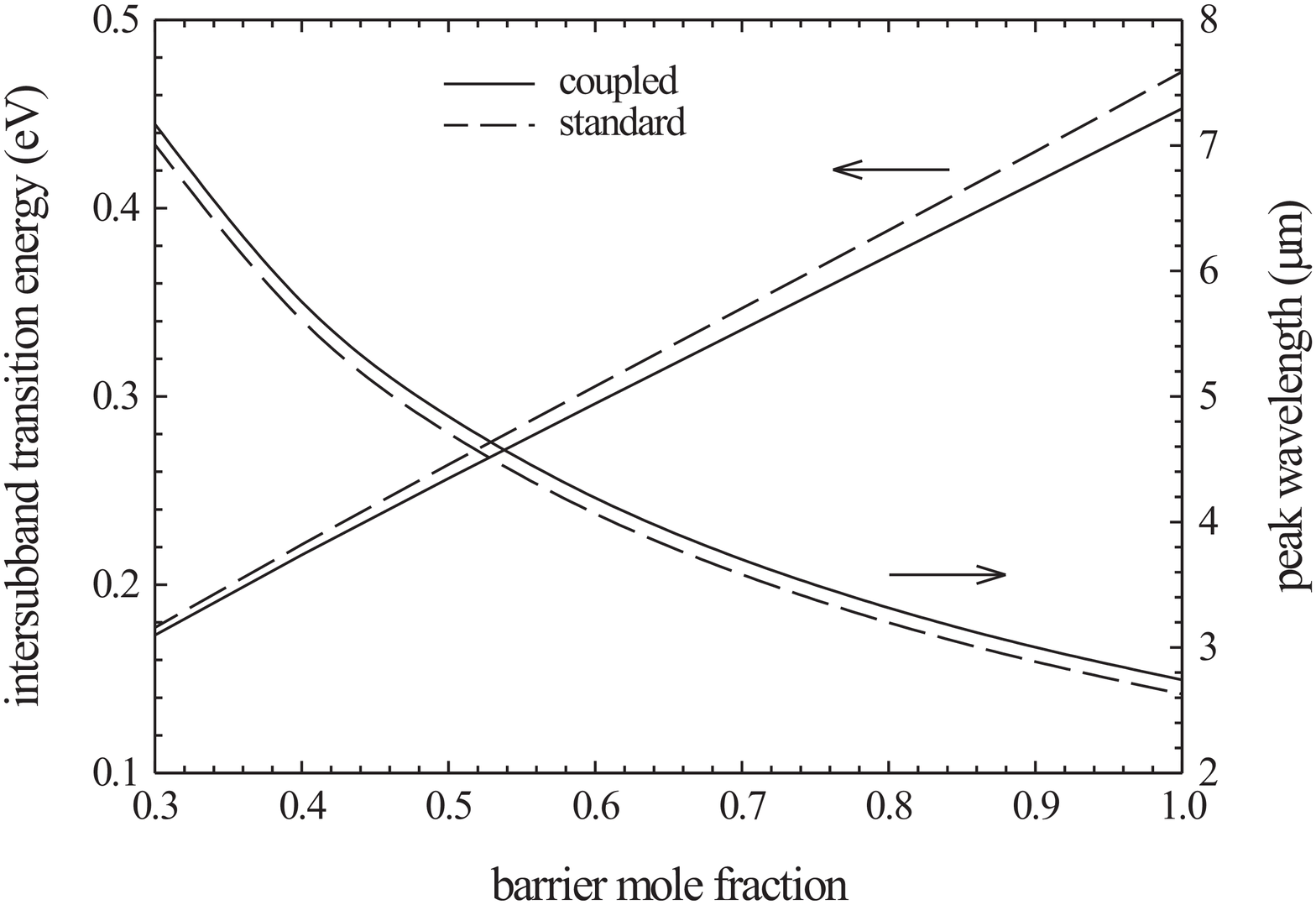}
  \end{center}
  \caption{Calculated ISBT energies (left $y$ axis) and peak
    wavelengths (right $y$ axis) between the first and second subbands
    at $k_z$$(w_a$$+$$w_b)$$=$$\pi$ for the SL described in Fig.\
    \protect\ref{fig:ezz} as a function of Al mole fraction $x$ in the barrier.  The
    fully coupled (solid line) and standard (dashed line) results are
    shown.}
  \label{fig:energies}
\end{figure}

Figure \ref{fig:energies} shows the calculated ISBT energies
between the first two electron subbands and the corresponding peak
wavelengths for the model SL.  The energies are calculated at
$k_z$$(w_a$$+$$w_b)$$=$$\pi$, the location of the minimum energy
separation between the first two subbands in the Brillouin zone.
The present calculations show that there is little change in the
energies between the zone center and zone boundary for a wide
range of SLs.  A number of factors contribute to the relatively
narrow mini-bandwidth.  First, the band edge discontinuity $
\Delta E_c $ is quite large due to the large band gaps of the host
materials.  Second, the effective electron mass is large, in this
case, $ 0.2 m_0 $ in GaN and $ 0.33 m_0 $ in AlN. Third, the
built-in electric field causes the electron wave function to be
localized in the triangular notch close to the AlGaN/GaN interface
(see Fig.\ \ref{fig:superlat}).  All of these factors reduce the
exponential tail of the electron wave functions between adjacent
wells, which, in turn, would appear as a dispersion in the mini
Brillouin zone.  For some SLs, however, particularly those with
thin wells, the wave functions will spread into the barrier
layers, causing some dispersion in the Brillouin zone.

The most significant feature of Fig.\ \ref{fig:energies} is the
discrepancy between the standard and fully-coupled models.  For
example, for $ x = 0.3 $, the fully-coupled transition energy is
lower than the standard value by 3.7 meV.  This difference
increases to 19.5 meV for $ x = 1 $.  The latter result is
especially significant, because high Al fractions are preferred
for optical switching technology due to the shorter peak
wavelength. The difference in energies between the two models is
large enough to be measurable by, for example, Fourier-transform
infrared spectroscopy (FTIR).  The wavelength for the standard
model is shorter by about 4\% relative to the coupled model for $
x = 1 $.

The reason for the red shift of the fully-coupled results relative
to the standard model can be understood by noting that the
introduction of electromechanical coupling reduces the magnitude
of the electric field (see Fig.\ \ref{fig:efield}). The smaller
electric field results in a conduction band profile closer to
flat-band conditions (see Fig.\ \ref{fig:superlat}). In addition,
a lowering of the effective barrier height occurs from the
reduction of $ \gamma_{zz}^\textrm{a} $ seen in Fig.\
\ref{fig:ezz}.  The more shallow triangular notch, together with a
reduced barrier height, will cause the subbands to more closely
spaced in energy. The calculated red shift of the ISBT is distinct
from the Stark shift seen in \textit{interband} transitions where
the transition energy shifts to \textit{higher} energy as the
electric field is reduced.

\subsection{Doped Superlattices}
\label{subsec:dopedSL} For optical switching technology, it is
necessary to $n$-dope the SL in order to populate the first
electron subband to facilitate ISBTs. For such structures, we use
the fully-coupled numerical model described previously.  The model
is tested against published optical data for various SL
structures. Table \ref{tab:comparison} shows the calculated ISBT
energy between the first two subbands for ten SL samples taken
from the literature.  The standard and fully-coupled results are
contrasted.  It is evident that two models give differing results.
Also evident is the consistent red shift of the fully-coupled
results compared to the standard results for the reasons discussed
in Sec.\ \ref{subsec:undopedSL}.  The differences depend on the
layer thicknesses and doping of the samples, varying from 4.9 meV
for sample A to 19.4 meV for sample D.  These differences are
significant enough to be measurable by standard techniques such as
FTIR.

Also shown in Table \ref{tab:comparison} are the
experimentally-obtained peak wavelengths for the SLs. These are
compared with the calculated wavelengths from the standard and
fully-coupled models.  Except for samples F and G, it is clear
that the calculated wavelengths are in reasonably good agreement
with the published data.  The causes of the discrepancies for
samples F and G are unclear at this point.  It should be noted
that we have not attempted to optimize the input parameters and
have chosen instead to use a generic set of
parameters\cite{hfetpara} without fitting. The calculated results
are very sensitive to all of the input parameters and also to the
geometry and Al fraction. For instance, if the well thicknesses in
samples F and G are increased by two monolayers and $x$ reduced to
0.6, the wavelength can be fitted to within 5\% using the
fully-coupled model.  More precise modeling of optical data will
be the subject of future work.  For now, we simply wish to
illustrate the importance of incorporating a fully-coupled strain
model in the design of optical switches.

The calculations in Table \ref{tab:comparison} were done for a
temperature of 300K.  At 77K, there is a blue shift of the
transition energies due to the slight increase in $ \Delta E_c $.
The blue shift is largest for SLs with the thinnest wells wherein
the subbands are pushed closer to band edge discontinuity and
smallest for SLs with the thickest wells in which the first two
subbands see less of the band edge discontinuity.  For example,
the shift is about 8.5 meV for sample A and about 0.15 meV for
sample E.

\begin{figure}
  \begin{center}
    \includegraphics[width=3.2in]{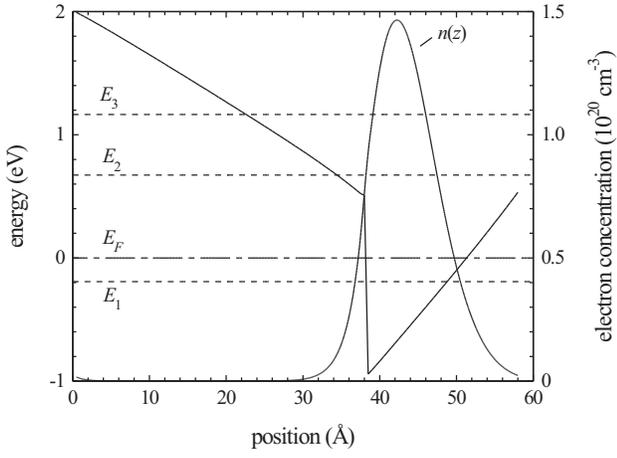}
  \end{center}
  \caption{Calculated conduction band edge (left $y$ axis)  and
    electron distribution (right $y$ axis) in one SL period for sample C
    in Table \protect\ref{tab:comparison} using the fully-coupled
    numerical model.  The first three electron subbands (dashed lines),
    calculated at $k_z$$(w_a$$+$$w_b)$$=$$\pi$, and the Fermi energy
    $E_F$ (dot-dashed line) are shown.}
  \label{fig:bandedge}
\end{figure}

Figure \ref{fig:bandedge} shows the calculated conduction band
edge and electron distribution for sample C in Table
\ref{tab:comparison} using the fully-coupled model.  Also shown
are the Fermi energy and the first three electron subbands
calculated at the Brillouin zone boundary.  This profile was
calculated at 300K.  At 77K, there is no discernible change in the
electron distribution function and the slope of the conduction
band edge.  There are, however, shifts in the subbands of a few
meV depending on the structure, as described earlier.  As the
calculation shows, the Fermi energy appears slightly above the
first subband but well below the second subband, in spite of the
high doping, ensuring that the first subband is populated by
electrons and the second nearly empty in order to facilitate
optical absorption.  This Fermi energy position is consistent with
measured SL structures with transition energies corresponding to
$E_1$$\rightarrow$$E_2$ transitions. The calculated distribution
and band edges, therefore, appear plausible.

\begin{figure}
  \begin{center}
    \includegraphics[width=3.2in]{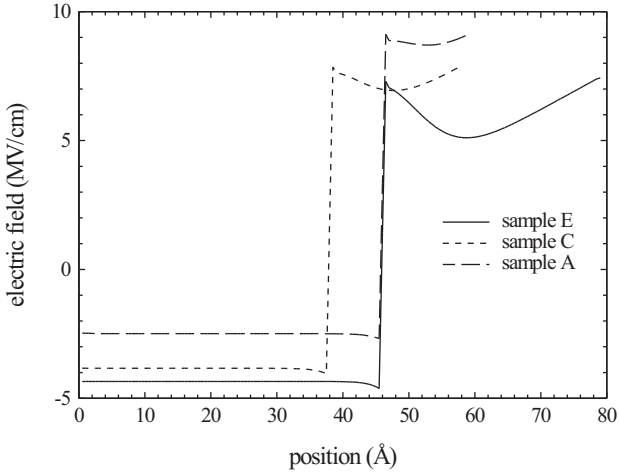}
  \end{center}
  \caption{Calculated electric field in one SL period for samples A, C,
    and E in Table \protect\ref{tab:comparison} using the
    fully-coupled numerical model.  The sign change in the electric
    field marks the position of the AlGaN/GaN interface.}
  \label{fig:fielddistrib}
\end{figure}

Figure \ref{fig:fielddistrib} shows the electric field
distribution for selected structures from Table
\ref{tab:comparison} using the fully-coupled model.  The large
electric fields in these structures are a consequence of the large
polarization discontinuity across the interface.  It is difficult
to verify these fields directly.  There is indirect evidence,
however, that these fields are not unreasonable given the close
fits of the ISBT wavelengths with experimental data.  Due to the
heavy doping, analytical expressions commonly used to estimate the
electric fields would lead to errors, especially in the wells
where the field is clearly non-linear.  Even on the barrier side
near the interface, there is an in increase in the magnitude of
the field due to the penetration of the wave functions into the
barrier. For such SLs, a numerical solution of the fully-coupled
Poisson equation as describe here is essential.

\section{Summary and conclusions}
\label{sec:summary} In summary, a fully-coupled model for the
strain and the eigenstates of AlGaN/GaN polarization SLs has been
presented.  This model is compared with the standard strain model
utilizing Hooke's law.  Both the spontaneous and piezoelectric
polarizations are included, together with free electrons and ionic
space charges.  It is seen that the strain and electronic
properties of the material are linked through the fully-coupled
thermodynamic equation of state for piezoelectric materials.
Separating the mechanical and electronic aspects of the SL in any
theoretical modeling of the properties of these structures leads
to errors in both the strain and the eigenstates of the system.
For strongly coupled cases, such as AlGaN/GaN SLs, the corrections
to the standard model can be significant. The ISBT energies
calculated from the fully-coupled model show a measurable red
shift compared to the corresponding energies calculated from the
separable model. This result has consequences for the design of
optical switches utilizing AlGaN/GaN SLs.

\acknowledgments The work of BJ was partially supported by the Air
Force Office of Scientific Research (AFOSR) and performed at Air Force
Research Laboratory, Materials and Manufacturing Directorate
(AFRL/MLP), Wright Patterson Air Force Base under USAF Contract No.\
F33615-00-C-5402.

\bibliography{supercoupled}

\begin{table}
  \caption{Comparison of published intersubband data with our
    calculated results for $ \textrm{Al}_x \textrm{Ga}_{1-x} $N/GaN SLs.
    The calculated energies in eV represent the separation between the
    first and second subbands at $k_z$$(w_a$$+$$w_b)$$=$$\pi$.  The peak
    wavelength $ \lambda $ is in $\mu$m.  The experimental (exp)
    results are contrasted with calculated results from the standard
    (std) and fully-coupled (cpl) models.}
  \label{tab:comparison}
  \begin{ruledtabular}
    \begin{tabular}{lccccccc}
      SL & $ w_a / w_b $ (\AA{}) & $x$ & $ E_{1 \rightarrow
      2}^\textrm{std} $ & $ E_{1 \rightarrow 2}^\textrm{cpl} $ & $
      \lambda^\textrm{exp} $ &
      $ \lambda^\textrm{std} $ & $ \lambda^\textrm{cpl} $  \\ \hline
      A\footnotemark[1] & 46/13 & 1 & 0.9521 & 0.9472 & 1.33 & 1.30 & 1.31 \\
      B\footnotemark[1] & 46/18 & 1 & 0.9040 & 0.8930 & 1.48 & 1.37 & 1.39 \\
      C\footnotemark[1] & 38/20 & 1 & 0.8610 & 0.8467 & 1.6 & 1.44 & 1.47 \\
      D\footnotemark[1] & 46/33 & 1 & 0.7057 & 0.6863 & 1.85 & 1.76 & 1.81 \\
      E\footnotemark[1] & 46/45 & 1 & 0.5800 & 0.5613 & 2.17 & 2.14 & 2.21 \\
      F\footnotemark[2] & 30/30 & 0.65 & 0.4981 & 0.4879 & 3 & 2.49 & 2.55 \\
      G\footnotemark[2] & 30/60 & 0.65 & 0.3740 & 0.3630 & 4 & 3.32 &
      3.42 \\
      H\footnotemark[3] & 27.3/13.8 & 1 & 0.9809 & 0.9718 & 1.27 &
      1.27 & 1.28 \\
      I\footnotemark[3] & 27.3/16.1 & 1 & 0.9352 & 0.9240 & 1.37 & 1.33
      & 1.34 \\
      J\footnotemark[3] & 27.3/22.6 & 1 & 0.8052 & 0.7874 & 1.54 &
      1.54 & 1.58 \\
    \end{tabular}
  \end{ruledtabular}
  \footnotetext[1]{Ref.\ \onlinecite{Iizuka2}, wells Si-doped $ 8 \times
    10^{19} $ $ \textrm{cm}^{-3} $.}
  \footnotetext[2]{Ref.\ \onlinecite{NSuzuki3}, wells Si-doped $ 4
    \times 10^{18} $ $ \textrm{cm}^{-3} $.}
  \footnotetext[3]{Ref.\ \onlinecite{Kishino}, wells Si-doped $ 10^{19}
    $ $ \textrm{cm}^{-3} $.}
\end{table}

\end{document}